\UseRawInputEncoding

\documentclass[12pt]{article}
\usepackage{blois,graphicx}

\begin{document}

\title{Triple-Pomeron Diffraction Peak as a Signature of Ultra-High-Energy Proton-Initiated Spectra of Gammas and Neutrinos in Astrophysics}

\author{O.I. Piskounova}

\address{P.N.Lebedev Physical Institute of Russian Academy of Science, Leninski prosp. 53, 119991 Moscow, Russia}



\maketitle\abstract{Production of ultra-high-energy (UHE) particles in astrophysics should not be different from the hadron production at the contemporary proton colliders. LHC experiments are providing us with the proton spectra at high energy (HE) that are measured in center-of-mass system (c.m.s.). The proton spectrum in c.m.s. has two special features that are important at HE hadron collisions: the growing central-rapidity part with hadron-antihadron pair production and the valuable diffractive peak in the region of leading proton due to the triple-pomeron term. The QCD phenomenological studies of previous years gave us the Quark-Gluon String Model for the calculations of baryon and meson production spectra in full kinematical range of rapidities. The collider proton production spectrum was recalculated to the laboratory system of coordinates, which is natural for the astrophysical observations. The specifics of collider distributions of protons are reflected in astrophysical spectra as the knee of cosmic proton spectrum and as the spectrum bulge at the UHE area  in cosmic spectra of protons, neutrinos, and gamma-photons. They help us to conclude that the energy distribution of baryon production plays an important role in the production of UHE cosmic particles. The remarkable result of this approach is the estimation of maximal energy ($E_{max}$) of initial protons (and antiprotons) from the energy of knee in proton spectrum that give us $E_{max} = 6. * 10^{12}$ GeV.}

\pagestyle{plain}


\section{Introduction}

Since it is not yet clear what part is taken by baryons in the dynamics of mass/energy transfer in space, the aim of this paper is to conclude about the influence on the forms of cosmic particle spectra \cite{partdata} by the specifics of HE collider proton production spectrum. As it has been shown previously \cite{piskoun20} on the basis of LHC data, the form of produced spectra can not change because of the  changes in hadronic interaction up to the c.m.s. energy $\sqrt{s}$ = 7 TeV that corresponds to the energy $E_{lab}$ = $2.5 * 10^7$ GeV, which is higher than the energy of knee in cosmic proton spectrum. Here, I suggest that the spectra of hadron production should not change their forms up to the highest available energy of  astrophysical protons. As the standard secondary proton spectra there will be taken the form of distribution from the Quark-Gluon String Model calculations \cite{kaidalov86} for the proton-proton collision at $\sqrt{s}$ = 540 GeV in c.m.s., which corresponds to the energy $1.6 * 10^{5}$ GeV in the laboratory system. The theoretical aspects of the proton spectrum features, which are surviving at UHE, are announced in Section 1.
 
I am going to study the form of spectra of cosmic particles after the first collision of protons that are injected from the UHE particle source in the space. The method of tranfer the collider spectra to laboratory system is shown in Section 2.

The form of resulting proton spectrum in laboratory system is described in Section 3. 

The comparison and discussion of experimental astrophysical neutrino and gamma distributions is done in Section 4. 

\section{QGSM proton spectrum at the c.m.s. proton collision with the energy $\sqrt{s}$ = 540 GeV}

First of all let me briefly describe the theoretical basis of HE hadroproduction. From the point of view of hadron physics, the pomeron-exchange regime is dominating in the production of HE particles as the energy of colliders entered in new higher ranges. The pomeron-based scheme of hadron collisions differs from the dynamics of low energy hadronic processes with quark-antiquark annihilation. These processes have been comprehended with huge theoretical and phenomenological efforts between the endings of sixties of last century \cite{pomeranchuk,chen,veneziano} and the beginnings of nineties \cite{kaidalov82,KTM,Cygnus90}. It has been understood that main contributions to HE baryon collisions are brought by multi pomeron exchanges. This leads to the growing total cross sections as well as to unification of inelastic cross sections of proton-proton and proton-antiproton collisions. Since the annihilation of q and $\bar{q}$ becomes impossible at high energies, the quark-gluon diagrams go to three dimensions and are generalized as multi pomeron exchanges with the contribution of diffraction scattering to the forward region of proton production spectra.

The achievements of long-time studies have been incorporated into the Quark-Gluon String Model.    
The spectra of protons from the collider data were described in QGSM in the middle of eighties of last century \cite{VKT85,kaidalov86,myPHD}. The many works have been done by the group of theorists \cite{arakelyan} in order to confirm and to supplement our calculations.

The growing density of particles at the central rapidities was already known. This is vacuum pairs production area, where
the production densities for protons and antiprotons are identical. The peak of proton production at the end of spectra was discussed already in \cite{pomeranchuk,chan,kaidalov79}. It is caused by the dominating contribution of the triple-connected vacuum trajectories to the elastic scattering, which influences on the inelastic proton production in the diffraction scattering too. At that time, the intercept of pomeron trajectory was asumed  equal to 1 and production cross sections did not grow yet, it was written in \cite{pomeranchuk} that "The triple-Pomeron term is the only scaling term, which increases as x close to 1."

\begin{figure}[h]
\centering
\includegraphics[width=12.0cm, angle=0]{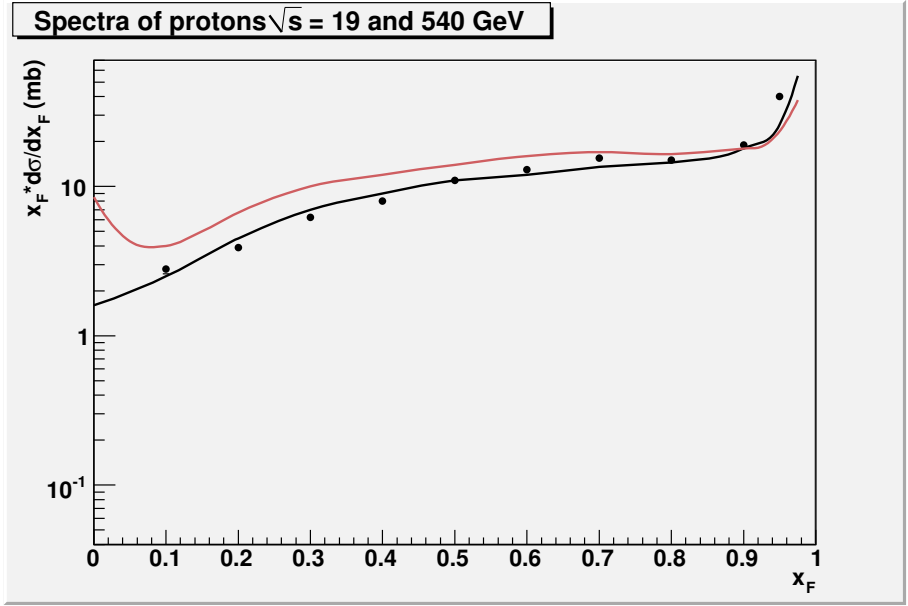}
 \caption{QGSM production spectra of protons in p-p collisions at colliders: the prediction for proton spectrum at the energy $\sqrt{s}$ = 540 GeV is shown with red line , QGSM proton distribution at lower energy, $p_L = 175 GeV/c$, is shown with solid line \protect\cite{kaidalov86}.}
 \label{proton540}
\end{figure}

\section{Transfer to the laboratory system}

The procedure for recalculating of c.m.s. spectrum into the laboratory system of coordinates is easy, it is shown in Figure 2. This method was applied in the early description of gamma spectrum from supernova \cite{Cygnus90}. The LHC experiments observed the central rapidity plateau in the spectra of produced hadrons. This spectrum should be a) shifted to the right by the value of $y_{max}$ = ln(2$\sqrt{s}$/$m_p$), and b) divided per $E_{lab}$.
 
\begin{figure}[h]
\centering
\includegraphics[width=12.0cm, angle=0]{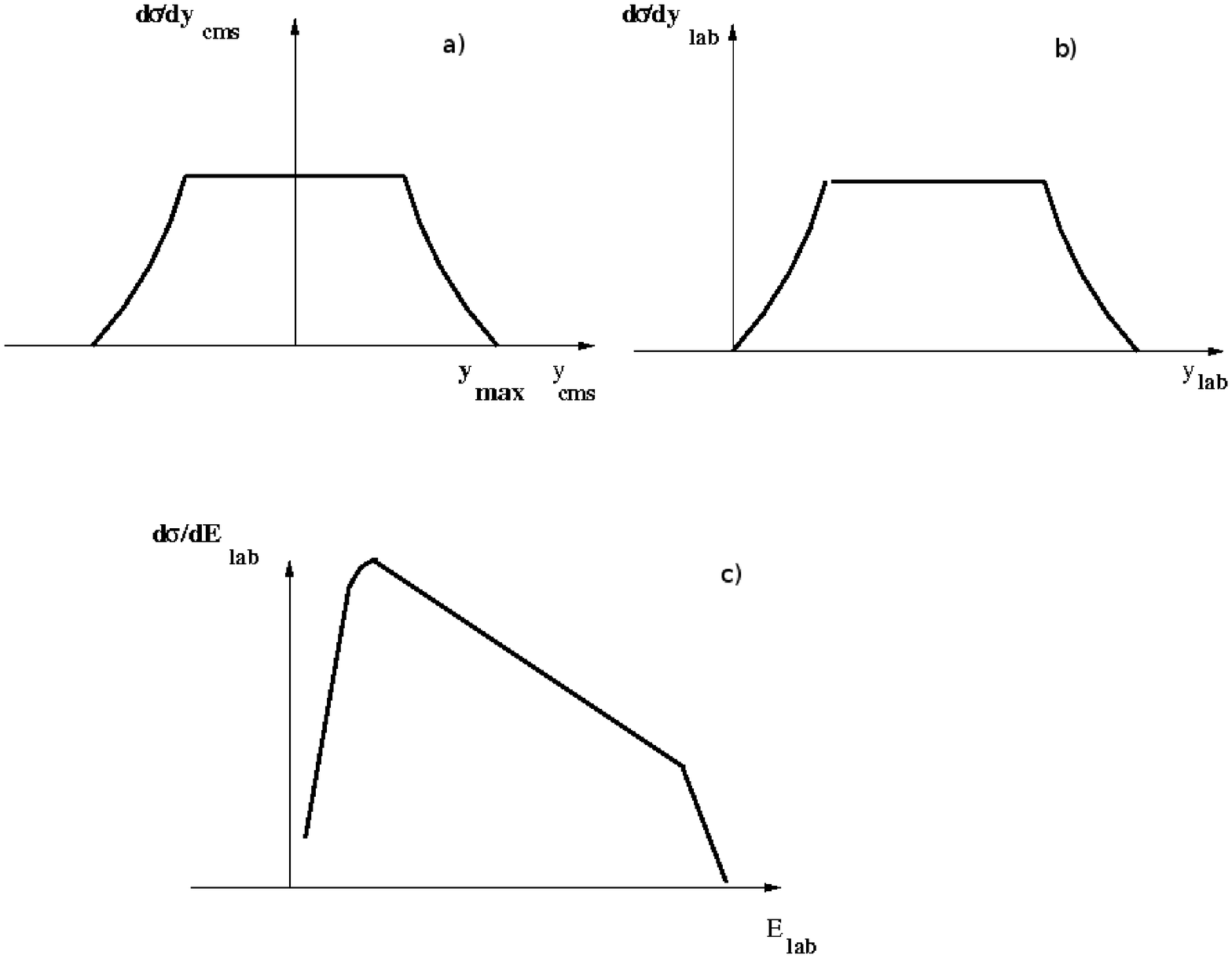}
 \caption{The procedure of c.m.s. spectrum transfer to the energy distribution of particle production in laboratory system.}
 \label{method}
\end{figure}
The procedure of spectrum modification is shown in Fig.~\ref{method}, where the c.m.s. distribution a) is symmetrical about negative and positive rapidities and has the shape of the plateau (or a bell) arround $ y_{cms}$=0. This spectrum should be b) shifted to the right by the value of $y_{max}$ = ln(2$\sqrt{s}$/$m_p$), and c) divided per $E_{lab}$.
 In such a way, the energy distributions in laboratory system have automatically the slope $E_{lab}^{-1}$.

\begin{equation}
\frac{d\sigma}{dE_{lab}} = E_{lab}^{-1}*\frac{d\sigma}{dy_{lab}}
\label{transfer}
\end{equation}
 
This method was applied in the reference \cite{Cygnus90} for the analysis of the spectrum slope for the gamma radiation from supernova.

\section{Resulting proton spectrum in space}

\begin{figure}[ht!]
\centering
\includegraphics[width=12.0cm, angle=0]{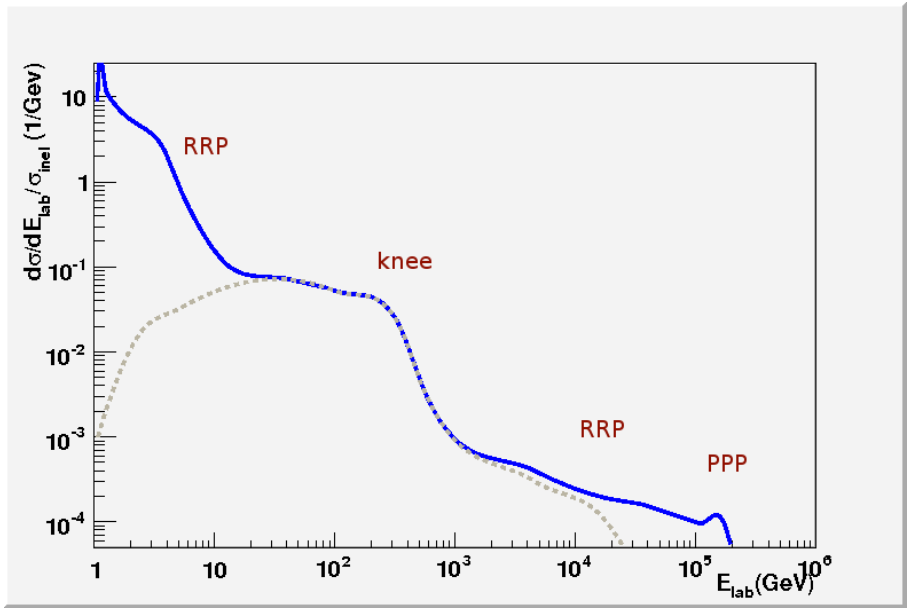}
 \caption{The form of proton spectra after UHE p-p collision in laboratory system, the antiproton spectrum is shown with dashed line. The contribution from triple-pomeron term (PPP) is seen at the end of spectra, the contribution from diquark transfer (RRP) should degenerate at UHE energy.}
 \label{spaceprotons}
\end{figure}

The form of proton energy spectrum in laboratory system is shown in the figure~\ref{spaceprotons}. In laboratory system, the central rapidity plateau gives us the knee and the triple-Pomeron peak corresponds to the bump at the end of Cosmic-Ray proton spectrum. At the same time, the triple-pomeron peak exists in antiproton spectrum too, if the UHE source injects the same amount of antiprotons as protons. In the other words, the specifics of HE leading spectra of protons and antiprotons do not contribute into the baryon asymmetry. At the same time, antiprotons should be absorbed by the positive-baryon environment of space.

It should be noticed too that if we are taking the position of knee in the real cosmic proton spectrum, we can estimate the maximal energy of primary proton from the UHE source, because the end of spectrum correspons to maximal rapidity in laboratory system that is a doubled maximal rapidity for the c.m.s.. As the result of taking into account the energy of knee in experimental proton spectrum \cite{partdata}, $E_{max}$ has be of order $E_{max} = 6. * 10^{12}$ GeV. 

\section{Spectra of neutrinos and gamma-photons}

\begin{figure}[h]
\centering
\includegraphics[width=14.0cm, angle=0]{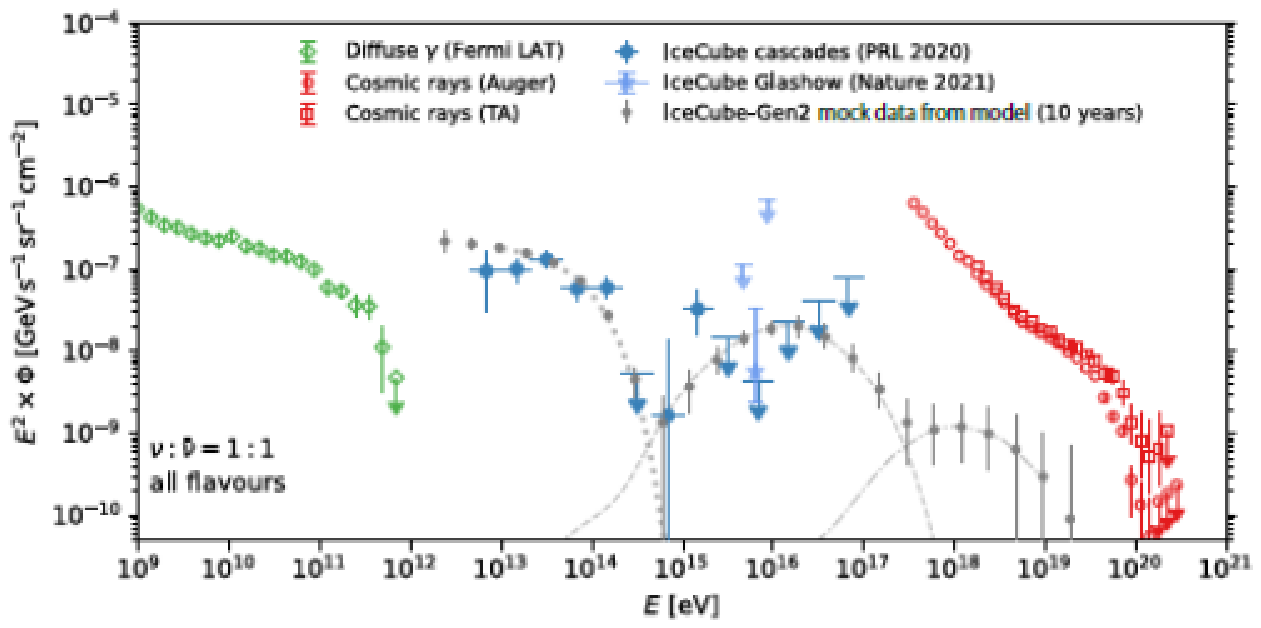}
 \caption{Spectra of different yelds in cosmic rays from the Snowmass review \protect\cite{snowmass}.}
 \label{neutrino}
\end{figure}

The buldge at highest energies is also seen in the spectrum of UHE neutrinos \cite{snowmass} in the figure~\ref{neutrino}. We know that neutrino appears from the decays of heavy baryons that is why the energies of neutrino spectrum are closer to initial proton energy. 
The gamma spectrum in the entire energy diapason, which was presented on the front cover of the book \cite{gammabook} also has an explicit bump at the end of distribution as well as the gamma distribution that is shown in another compilation \cite{gzk}. It is the clear that both spectra bring the signature of UHE proton interaction.


It is important to note that no any other mechanism is known to enhance the particle production in this very energetic region of distributions. The well known GZK effect should on the contrary just cut spectrum in this area \cite{gzk}.

As a working hypothesis, I have a guess that DM is built in some symmetric way, from protons and antiprotons, which are compressed under extreme gravitation pressure in the Super Massive Black Holes (SMBH). The jet activity from SMBH is anything, but the catastrophe of massive gravitational objects. It seems, the periodical giant jets of Baryonium Dark Matter \cite{TorusDM} take place due to the nonequilibrium prosesses inside the SMBH. 

\section{Conclusions}

Protons and antiprotons are the fundamental matter in the Universe. They can not just annihilate in HE collision that will  the baryon matter in space. Their interaction at high energies is the source of many other particles in cosmic rays: secondary protons and antiprotons, positrons and electrons, gamma-photons, and neutrinos \cite{poster,uhecr}.

The UHE protons and antiprotons are injected with jets from SMBHs, which are the most powerful sources in the Universe, and initiate  particle spectra with distinctive signatures. If injected symmetrically towards baryon charge, the spectra of HE antiprotons from antiproton-antiproton interaction should have the same features as in proton-proton case. The production processes do not bring additional baryon asymmetry. Might be that antiprotons are absorbed stronger than protons in the mostly positive-charged baryon environment of space.

Proton spectrum in the laboratory system has the similar features as the energy distribution at the proton accelerator experiments. If transfered into laboratory system, the central rapidity plateau is converted into the knee of laboratory spectrum, and the triple-pomeron peak corresponds to the buldge at the end of spectra of cosmic particles: protons, neutrinos, gamma, and others. It should be taken into account that every additional hadronic collision (or decay) brings the factor $E^{-1}$ into the resulting energy spectrum of secondary particles.

In this approach we can estimate the initial energy of protons (antiprotons) from the source. Since the knee corresponds to rapidity center in c.m.s., the maximal rapidity in astrophysical proton spectrum is larger than the rapidity of knee by a factor 2. In such a way, the maximal energy of spectra is approximatelly $E_{max} = 6. * 10^{12}$ GeV that corresponds to $6. * 10^6$ PeV or 6. ZeV.

\section{Suggestions for futher studies and experiments}

First of all, the proton production spectra at highest collider energy have to be measured in the diffraction region, $x_F$ close to 1, in order to approve the valuable contribution of triple-pomeron diffraction term. The  phenomenological calculation adjusted for HE proton spectrum has to follow the experimental measurements and, in addition, to show how the contribution from diquarks in the middle area of $x_F$ degenerates with the growing of energy. It will be nesessary to compair the form of UHE proton spectra with the energy distribution of UHE cosmic neutrinos. The comparison of experimental spectra of neutrino and antineutrino in the UHE region, if possible, will show whether an equal production exists for UHE protons and antiprotons from SMBHs.  

\section{Acknowledgements}

Author expresses her gratitude to Prof. Oleg Kancheli for the numerous advices towards HE hadroproduction, diffractive reggeon processes and cosmic ray spectra.


\section{References}

\begin{thebibliography}{99}
\bibitem{partdata}M.Tanabashi et al. {\it Review of Particle Data Group}, Phys. Rev. D 98 (2018) 030001
\bibitem{piskoun20}O.I.Piskounova, {\it Baryon production at LHC experiments: average pt of hyperons versus energy}, Int. Jou. of Mod. Phys. A {\bf 35} (2020) 2050067 [arXiv:1706.07648].
\bibitem{pomeranchuk}V.N.Gribov, I.Ya.Pomeranchuk and K.A.Ter-Martirosyan, {\it Partial Waves Singularities Near j=1 and High Energy Behaviour of the Elastic Scattering Amplitudes}, Phys. Lett. {\bf 9} (1964) 269.
\bibitem{chen}M.-S.Chen et al., {\it Phenomenological Study of Single Particle Distributions near the Kinematical Limits}, Phys. Rev. D{\bf 7} (1972) 1667.
\bibitem{veneziano}G.Veneziano,{\it Origin and intercept of the Pomeranchuk singularity}, Phys Lett. B {\bf 43} (1973) 413.
\bibitem{chan}H.-M.Chan, J.F.Paton and S.T.Tsou, {\it Diffractive Scattering in the Dual Model}, Nucl. Phys., B{\bf 86} (1975) 479.
\bibitem{kaidalov79}A.B.Kaidalov, {\it Diffractive Production Mechanizms}, Phys. Rep. {\bf 50} (1979) 159.
\bibitem{kaidalov82}A.B.Kaidalov, {\it The Quark-Gluon Structure of the Pomeron and the Rise of Inclusive Spectra at High Energies}, Phys. Lett. B{\bf 116} (1982) 459. 
\bibitem{KTM}A.B.Kaidalov and K.A.Ter-Martirosyan, {\it Multiple Production of Hadrons at High Energies in the Model of Quark Gluon Strings. Theory.}, Sov. Journal of Nucl. Phys. {\bf 39} (1984) 979,{\it ibid} {\it Comparison with Experiment.} {\bf 40} (1984) 135.
\bibitem{VKT85}A.I.Veselov, O.I.Piskunova, K.A.Ter-Martirosian, {\it Production and Decay of Quark-Gluon Strings: Inclusive Distributions of Hadrons in $p_t$ and x}, Phys. Lett.B {\bf 158} (1985) 175. 
\bibitem{kaidalov86}A.B.Kaidalov and O.I.Piskunova, {\it Inclusive Spectra of Baryons in the Quark-Gluon Strings Model}, Z. Phys. C{\bf 30} (1986) 145.
\bibitem{myPHD}Olga I.Piskunova, PhD thesis (in russian), 1989
\bibitem{Cygnus90}O.I.Piskunova, {\it Shape of Gamma Spectra from Cosmic Sources of High Energy Protons}, Sov. Journal of Nucl. Phys. {\bf 51} (1990) 1332.
\bibitem{arakelyan}G. H.Arakelyan, C.Merino, C.Pajares, Yu.M.Shabelski, {\it Midrapidity Production of Secondaries in pp Collisions at RHIC and LHC Energies in the Quark-Gluon String Model}, Eur. Phys. J. C{\bf 54} (2008) 577 [arXiv:0709.3174].  
\bibitem{snowmass}A.Coleman et al., {\it Ultra-High-Energy Cosmic Rays: The Intersection of the Cosmic and Energy Frontiers}, Astropart. Phys. {\bf 149} (2023) 102819 [arXiv:2205.05845].
\bibitem{gzk}M.S.Muzio, G.R.Farrar, and M.Unger, {\it Probing the environments surrounding ultrahigh energy cosmic ray accelerators and their implications for astrophysical neutrinos}, Phys. Rev, D{\bf 105} (2022) 023022 [arXiv:2108.05512].
\bibitem{poster}O.I.Piskounova, {\it Ultra-High Energy Proton-Proton Collision in the Laboratory System as the Source of Proton, Neutrino and Gamma Spectra in Astrophysics}, PoS ICHEP2022)(2022) 911 [arXiv:2110.13618].
\bibitem{uhecr}O.I.Piskounova, {\it Triple-Pomeron Diffraction Peak as a Signature of Proton Interaction at the
Sources of VHE Gamma and Neutrino in Space}, Presentation at UHECR2022, Gran Sasso, Italy, September (2022) [arXiv:2302.08546].
\bibitem{gammabook}F.A.Aharonian, {\it Very High Energy Cosmic Gamma Radiation, Crucial Window on the Extreme Universe}, World Scientific Publishing, 2004.
\bibitem{TorusDM}O.I.Piskounova, {\it High Energy Proton-Proton collisions and Baryonium Dark Matter}, LHEP {\bf 378} (2023) 1[arXive:2211.07649].

\end{thebibliography}
\end{document}